\documentclass[aps,prl,twocolumn,10pt,superscriptaddress,floatfix]{revtex4-2}

\usepackage{amsmath}
\usepackage{mathtools}
\usepackage{graphicx}
\usepackage{amsfonts}
\usepackage[colorlinks=true,allcolors=blue]{hyperref}
\graphicspath{{figures/}}

\begin{document}

\title{Yukawa screening derivation of the bond-valence rule}

\author{Michael L. Whittaker}
\email{mwhittaker@lbl.gov}
\affiliation{Energy Geosciences Division, Lawrence Berkeley National Laboratory, Berkeley, CA 94720}
\affiliation{Materials Sciences Division, Lawrence Berkeley National Laboratory, Berkeley, CA 94720}

\author{Pan Wang}
\affiliation{Energy Geosciences Division, Lawrence Berkeley National Laboratory, Berkeley, CA 94720}

\author{Chunhui Li}
\affiliation{Meta Platforms, Inc., Menlo Park, CA 94025}

\author{Naman Katyal}
\affiliation{Energy Geosciences Division, Lawrence Berkeley National Laboratory, Berkeley, CA 94720}

\author{Piotr Zarzycki}
\affiliation{Materials Sciences Division, Lawrence Berkeley National Laboratory, Berkeley, CA 94720}

\begin{abstract}
The bond-valence model is a standard way to estimate bond strengths in
crystals, but its exponential dependence on bond length has lacked a
derivation from a specific physical interaction.  We show that this form
emerges as the leading-order limit of screened Coulomb electrostatics
and that the fitted bond-valence softness can be interpreted in terms of
an electronic screening length.  This turns bond valence from an
empirical fitting rule into a transferable descriptor of local screened
charge response across coordination environments.  The resulting theory predicts how the
bond-valence parameters should vary with ionic charge and coordination
number, and that prediction agrees with 150 fitted valences from 94
cation--oxygen species, including 68 in fourfold coordination and 
82 in sixfold coordination, at an abundance-weighted
coefficient of determination of 0.986.  A comparison with
first-principles charge densities shows that the bond-valence shell 
radius tracks the electronic screening cloud with coefficients of determination 
of 0.9998 for ten alkali and alkaline-earth oxides and 0.967 for 21 other binary oxides
whose nearest-neighbor environments match the theory's assumptions.  The
widely used bond-valence model is thus the leading-order expression of
screened electrostatics in ionic solids.
\end{abstract}

\maketitle

Pauling's observation that the strength of a bond reaching a coordinating
anion approximates the cation charge divided by the coordination number
\cite{Pauling1929} has been an organizing principle of crystal chemistry
for nearly a century.  Brown and Altermatt \cite{Brown1985} cast this idea
as the bond-valence model: the empirical bond strength between a cation $i$
and anion $j$ separated by $R_{ij}$ is
\begin{equation}\label{eq:bv}
S_{ij}=\exp\!\left(\frac{R_0-R_{ij}}{B}\right),
\end{equation}
where $R_0$ and $B$ are parameters fitted to crystallographic data and the
valence-sum rule $\sum_j S_{ij}=z_i$ (formal charge) is enforced.  The
model is used routinely for structure validation
\cite{Brown2009,Brown2016,Gagn2015}.

Several lines of work have sought a physical foundation for the
bond-valence exponential.  Preiser \emph{et al.}\ \cite{Preiser1999} gave
a clear Gauss's-law account, interpreting bond valence as localized
electrostatic flux, and Brown developed this picture further within a
broader electrostatic framework \cite{Brown2009,Brown2016}. However, in those
treatments the exponential distance dependence is postulated rather than
derived from a specific potential, leaving $B$ as an empirical fit
parameter with length units but no associated physical grounding.  
Adams related the softness $B$ to Morse-potential curvature and motivated non-universal
$B$ values \cite{Adams2001,Chen2017}, but does not connect $B$ to a screening length.
Modern first-principles work can compute screened local interactions
from electronic structure \cite{Aryasetiawan2004,Shih2012}, but does not
derive the bond-valence form itself.  Here we show that a Yukawa
expansion obtains the exponential as the leading term of a specific
screened potential, ties $B$ to a measurable screening length, predicts
the $B$--$R_0$ correlation with no fitted constants, and connects the
fitted parameters to electronic structure.

\emph{Screened-flux derivation.} A cation with charge $Q_i=z_i e$ emits
electrostatic flux $Q_i/\epsilon_0$.  In a Yukawa field with screening
length $\lambda$, the signed flux amplitude captured by bond $i$--$j$ is
$\widetilde\Phi_{ij}\propto\Omega_{ij}\,e^{-R_{ij}/\lambda}(1+R_{ij}/\lambda)$,
where $\Omega_{ij}$ is the solid-angle fraction assigned to neighbor $j$.
Bond valence is a nonnegative partition of the cation valence among its
anion neighbors.  Imposing the valence-sum rule $\sum_j S_{ij}=z_i$ and
preservation of screened-flux magnitudes, $S_{ij}/S_{ik}=w_{ij}/w_{ik}$,
uniquely gives
\begin{equation*}
S_{ij}=z_i\,\frac{w_{ij}}{\sum_k w_{ik}},
\qquad w_{ij}:=|\widetilde\Phi_{ij}|.
\end{equation*}
Choose $R^*$ operationally as the Boltzmann-weighted mean bond length
$m_i:=\sum_j p_{ij}R_{ij}$ (defined fully in the partition-function
section below).  Expanding $\ln w_{ij}$ about $R^*$ gives
\begin{equation}\label{eq:expansion}
\ln w_{ij}= C_i - \frac{R^*}{\lambda(\lambda+R^*)}\,(R_{ij}-R^*)+\delta_{ij},
\end{equation}
where $C_i$ is independent of bond index $j$ and
$\delta_{ij}=O[(R_{ij}-R^*)^2/\{\lambda(\lambda+R^*)\}]$.
Exponentiating the linear
term and renormalizing therefore reproduces Eq.~\eqref{eq:bv} at
$R_{ij}=R^*$ with only quadratic log-weight error away from $R^*$.  Define
the leading-order Yukawa exponent denominator
\begin{equation}\label{eq:B_composite}
B = \frac{\lambda(\lambda+R^*)}{R^*}.
\end{equation}
This $B$ is the first-order exponent denominator produced by the Taylor
expansion of the Yukawa weight.  On the screening branch it is positive
and coincides with the conventional bond-valence parameter; on the
anti-screening branch it becomes negative. The screening length $\lambda$ is the primary physical
variable, while $B$ is its leading-order image in the bond-valence
exponential.  On the screening branch ($\lambda>0$, $B>0$), inverting
Eq.~\eqref{eq:B_composite}
at the shell level gives the structure-specific screening length
$\lambda=(-R^*+\sqrt{R^{*2}+4BR^*})/2$.  At the species-level
characteristic pair defined below, we evaluate the same positive-branch
inversion on the characteristic reference shell, for which the
operational choice $R^*=m_i$ gives $R^*=R_0^*$, so the tabulated
species constant is
$\lambda^*=(-R_0^*+\sqrt{R_0^{*2}+4B^*R_0^*})/2$ for species with
$B^*>0$.

\begin{figure}[tb]
\centering
\includegraphics[width=\columnwidth]{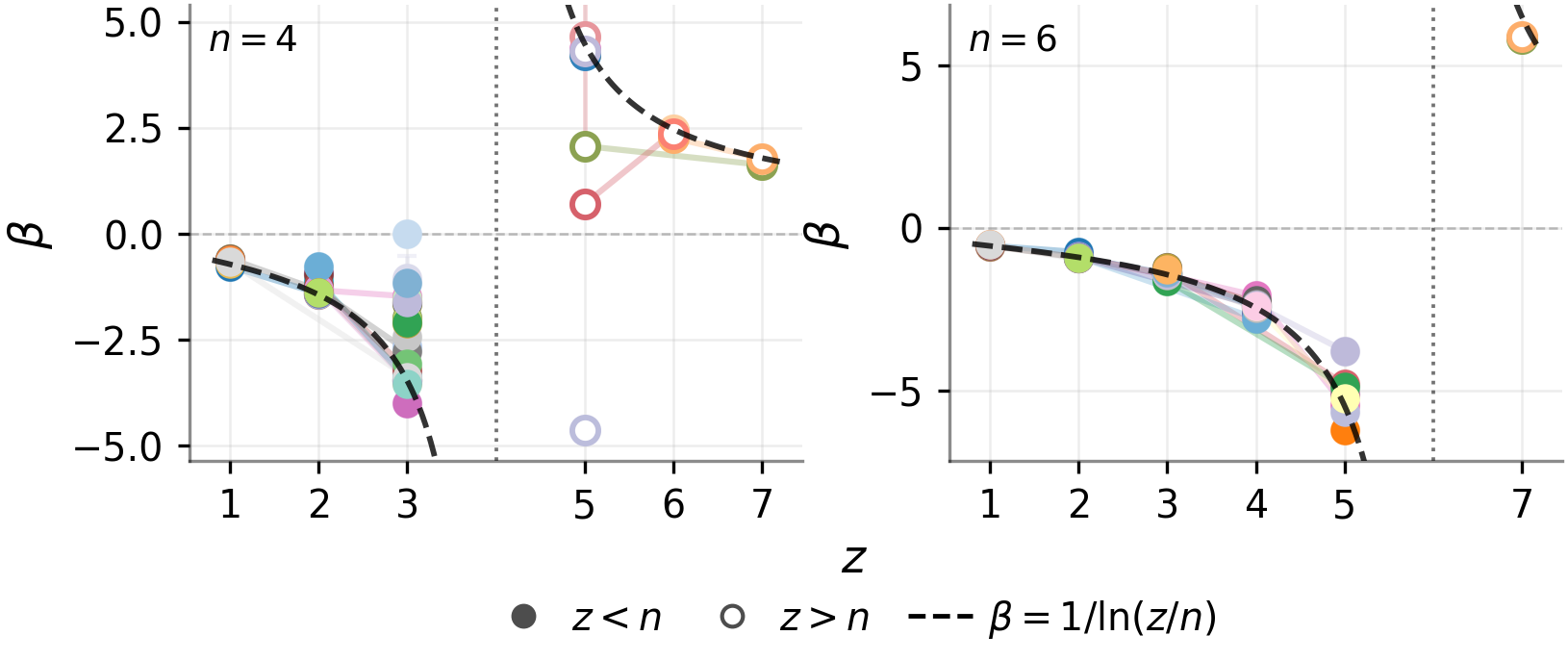}
\caption{Fitted slope $\beta$ vs.\ formal charge $z$ for 94
cation--oxygen species at $n=4$ (68 cations, left) and $n=6$ (82
cations, right); 56 cations contribute at both coordinations, giving
150 fitted slopes in total.  Dashed curves: $\beta=1/\ln(z/n)$.  The
pole at $z=n$ separates the screening ($z<n$) and anti-screening
($z>n$) branches; species at $z=n$ are omitted because the $B$--$R_0$
fit is degenerate at the pole (see Supplemental Material
\cite{suppmat}).}
\label{fig:pole}
\end{figure}

\emph{Parameter-free slope.} In the uniform-bond limit, where
$R_{ij}=\bar R$ for all $j$, the valence-sum rule gives
$R_0=\bar R+B\ln(z/n)$.  Differentiating:
\begin{equation}\label{eq:pole}
\beta := \frac{dB}{dR_0} = \frac{1}{\ln(z/n)}.
\end{equation}
This is parameter-free: the slope of the $B$--$R_0$ correlation at fixed
coordination number $n$ is determined entirely by the charge-to-coordination
ratio $z/n$, with a pole at $z=n$ marking the screening/anti-screening crossover.
Figure~\ref{fig:pole} shows that Eq.~\eqref{eq:pole} reproduces the
fitted slopes for 94 cation--oxygen species at $n=4$ and $n=6$ (68 at
$n=4$, 82 at $n=6$; 150 slopes in total; weighted $R^2=0.986$, mean
absolute error $0.15$; weighting scheme defined in the Supplemental
Material \cite{suppmat}).

\emph{Screening collapse at $z=n$.} The pole has direct physical
content.  At $z=n$ every bond carries exactly one valence unit
($S_{ij}=1$), so $\exp[(R_0-R_{ij})/B]=1$ for all~$j$.  For a
symmetric shell with $R_{ij}=\bar R$ this is trivially satisfied at any
$B$ with $R_0=\bar R$---the $z=n$ limit is degenerate.  For a real
coordination shell with unequal bond lengths, however, the only solution
is $B\to\infty$, so that every finite $(R_0-R_{ij})/B\to 0$.  On the
screening branch, Eq.~\eqref{eq:B_composite} gives
$\lambda\approx\sqrt{BR^*}\to\infty$: the screening length diverges.
On the $z>n$ side of the pole, $B$ changes sign ($B<0$).  Its magnitude
still diverges, so the branch change is carried directly by $\mathrm{sgn}\,B$
or, equivalently, by the sign reversal of $\beta=1/\ln(z/n)$.  Approaching the pole from either
side ($z=n\mp\varepsilon$, $\varepsilon\to 0$), the divergence rate is
governed by the species characteristic intercept $R_0^*$ (the
$n$-resolved-line convergence point introduced below):
\begin{equation*}
|B| \sim \frac{n\,|R_0-R_0^*|}{\varepsilon},
\qquad
|\lambda| \sim \sqrt{\frac{R^*\,n\,|R_0-R_0^*|}{\varepsilon}}.
\end{equation*}
The screening length diverges as $\varepsilon^{-1/2}$.  In conventional
screening theory, metals have finite Thomas--Fermi screening set by a
nonzero density of states at the Fermi level, whereas gapped or
correlated insulators suppress low-energy charge response and therefore
screen more weakly \cite{Thomas1927,Fermi1928,Debye1923,AshcroftMermin1976}.
Our coarse-grained $\lambda$ is not identical to those microscopic
lengths, but its pole at $z=n$ is consistent with the same weak-screening
limit: the bond-valence partition becomes insensitive to finite
bond-length differences and the fitted $B$--$R_0$ parameters cease to
resolve the shell.  Species that sit at or near $z=n$---Si$^{4+}$ in tetrahedral
coordination (SiO$_2$, band gap 9~eV), B$^{3+}$ in trigonal
coordination (B$_2$O$_3$, 6~eV), and C$^{4+}$ at $n=4$---lie in the
wide-gap covalent limit where such weak screening is physically
plausible.

\begin{figure*}[t]
\centering
\includegraphics[width=\textwidth]{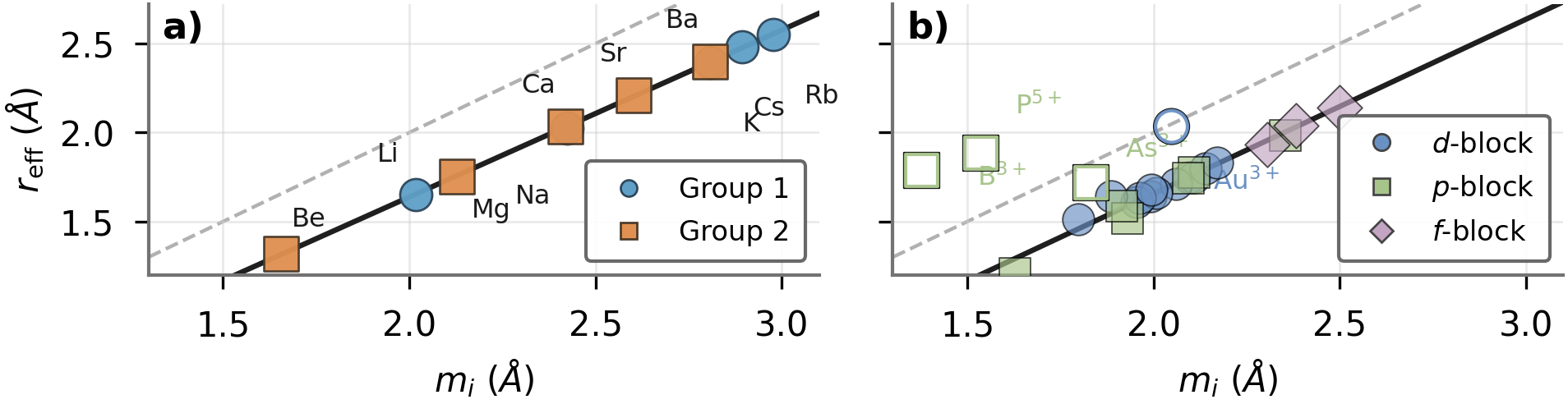}
\caption{First-principles screening centroid $r_{\mathrm{eff}}$ vs.\ Boltzmann shell
centroid $m_i$.  (a) Ten $s$-block ground-state binary oxides.  Solid
line: shared fit; dashed line: $r_{\mathrm{eff}}=m_i$.  (b)
Twenty-five non-$s$-block ground-state binary oxides, color-coded by
block.  Open symbols mark the four directional-bonding outliers
excluded from the regression (B$_2$O$_3$, P$_2$O$_5$, As$_2$O$_3$,
Au$_2$O$_3$).}
\label{fig:centroid}
\end{figure*}

\emph{Partition thermodynamics.} Because $R_0/B$ cancels in
$S_{ij}/\sum_k S_{ik}$, the normalized bond weights
$p_{ij}=S_{ij}/\sum_k S_{ik}=e^{-R_{ij}/B}/Z_i$ define a canonical
partition function with $B>0$ as the temperature-like scale on the
screening branch.  The Boltzmann
mean $m_i:=\sum_j p_{ij}R_{ij}$ and the Shannon entropy
$H_i:=-\sum_j p_{ij}\ln p_{ij}$ yield the exact identity
\begin{equation*}
R_0 = m_i + B(\ln q_i - H_i),
\end{equation*}
where $q_i:=\sum_j S_{ij}$.  Here $H_i$ is a purely geometric quantity:
it measures how uniformly the bond-valence weight is distributed across
the coordination shell, reaching $\ln n$ for a symmetric shell and
falling below it as distortion concentrates weight on the shorter bonds.
Thus $R_0$ is not the mean bond length but an entropy-corrected weighted
mean, and the operational choice $R^*=m_i$ in
Eq.~\eqref{eq:B_composite} is made explicit.

\emph{Empirical validation.} Bond-valence parameters $(R_0,B)$ were
determined for 103 cation--oxygen species using first-principles
relaxed crystal
structures from the Materials Project \cite{MaterialsProject,Horton2025}.
For each structure, bond strengths were obtained by solving the
valence-sum and loop constraints on the cation--oxygen bond graph
\cite{Li2025}, and $(R_0,B)$ were fitted by linear least squares; the
few rank-deficient bond graphs and numerically ill-conditioned fits
were treated by the procedure described in the Supplemental Material
\cite{suppmat,YukawaScreeningRepo}.  The resulting set spans $s$-,
$d$-, $f$-, and $p$-block cations; Fig.~\ref{fig:pole} shows the 94
species (150 fitted slopes) for which the prediction is finite at
$n=4$ and/or $n=6$.

Within each $(z,n)$ group, the slope $\beta^{(n)}$ confirms
Eq.~\eqref{eq:pole} across all families.  The $n$-resolved lines
converge to cation-specific characteristic pairs
$(R_0^*,B^*)$, defined as the intersection that minimizes
the weighted variance of the coordination-resolved $B$--$R_0$ lines,
with each line weighted by the final fit population of that
coordination shell (Supplemental Material \cite{suppmat}).
Of the 103 species, 101 have $B^*>0$ (screening branch);
for these the screening length
$\lambda^*=(-R_0^*+\sqrt{R_0^{*2}+4B^*R_0^*})/2$ is a tabulated
constant (Supplemental Material \cite{suppmat}).  The remaining two
species (Mo$^{3+}$, Sb$^{3+}$) have $B^*<0$, fall on the anti-screening branch, and are
retained only as signed diagnostics.  Both sit outside the first-order
Yukawa regime: Sb$^{3+}$ has a stereoactive $5s^2$ lone pair driving
strong directional distortion, and Mo$^{3+}$ is constrained by only two
coordination shells (Supplemental Material \cite{suppmat}), so its
$(R_0^*,B^*)$ is provisional.  The screening-branch $B^*$ values are non-universal
(mean $0.38\pm0.15$~\AA), in qualitative agreement with Adams's softBV
parameterization \cite{Adams2001}, while $\lambda^*$ provides a
single-parameter species descriptor.

\emph{First-principles confirmation.} The connection to electronic structure is tested
by comparing the bond-valence Boltzmann shell centroid
$m_i=\sum_j p_{ij}R_{ij}$ to a first-principles screening centroid
$r_{\mathrm{eff}}$ extracted from charge densities in the
Materials Project \cite{MaterialsProject,Horton2025}.  For each M--O
bond with $R_{ij}\le 3.5$~\AA, the total valence-electron density
$\rho(r)$ is sampled at 200 uniformly spaced points along the
nearest-image straight M--O segment.  An interior density-minimum proxy
on $0.15R_{ij}<r<0.85R_{ij}$ defines the start of the oxygen-side
window; $r_{\mathrm{eff}}$ is then the excess-density-weighted centroid
on $r_{\mathrm{BCP}}\le r\le 0.93R_{ij}$ with weights
$\max[\rho(r)-\rho_{\mathrm{BCP}},0]$, averaged over the valid
first-shell bonds of that oxide (see Supplemental Material for the exact
discrete construction and fallback rule).
For ten $s$-block binary oxides (Li$_2$O through BaO;
Fig.~\ref{fig:centroid}a),
these collapse onto a single line:
\begin{equation*}
r_{\mathrm{eff}} = 0.94\,m_i - 0.24~\text{\AA}
\qquad (R^2=0.9998).
\end{equation*}
This is not a trivial monotone size trend: the jackknife
mean absolute error of the $m_i$ regression is $0.006$~\AA, versus
$0.078$~\AA\ for a linear fit against $R_0$ and $0.095$~\AA\ against the
Shannon cation radius
\cite{Shannon1976}
(see Supplemental Material for the corresponding control plot).  The near-unit slope therefore shows
that the bond-valence partition function and the first-principles screening cloud track
the same ionic shell radius.  The same centroid correspondence extends
beyond the $s$ block.
Figure~\ref{fig:centroid}b shows the 25 non-$s$-block binary oxides for
which the charge density yields a well-defined oxide-level screening centroid
(listed in the Supplemental Material \cite{suppmat}).  The Yukawa
derivation assumes approximately isotropic first-shell screening, so
the regression is restricted to the 21 oxides whose cation environments
satisfy that condition, excluding four whose bonding is dominated by
directional, non-isotropic motifs: B$_2$O$_3$, P$_2$O$_5$,
As$_2$O$_3$, and Au$_2$O$_3$; the Supplemental Material gives the fixed
inclusion rule and the exact oxide-level centroid construction used to
define $r_{\mathrm{eff}}$.  The 21 included oxides follow
\begin{equation*}
r_{\mathrm{eff}} = 0.98\,m_i - 0.30~\text{\AA}
\qquad (R^2=0.967).
\end{equation*}
The centroid correspondence therefore extends across the $p$-, 
$d$-, and $f$-block with $R^2=0.967$, and its breakdown is confined to
strongly anisotropic directional-bonding environments outside the isotropic Yukawa regime.

\emph{Discussion.} The main result is that the bond-valence
exponential---the workhorse of crystal-chemistry validation for four
decades---is the leading-order term of Yukawa screened electrostatics.  The
fitted exponent denominator $B$ is not a free parameter but the
leading-order Yukawa denominator set by the screening length and shell geometry, with its sign carrying the screening/anti-screening branch information; the parameter correlation
$\beta=1/\ln(z/n)$ is parameter-free and exact in the uniform-bond limit;
and the partition-function structure gives an entropy-corrected identity
for $R_0$ and a free energy $F=m_i-BH_i=R_0-B\ln q_i$ in length units.

The first-principles centroid comparison confirms that this is not merely a
mathematical analogy: the electronic screening cloud and the bond-valence
weights describe the same shell geometry.  The small deviation from the
identity line is explained by the oxygen-side valence centroid shift
measured directly from the same charge-density paths, while the null-model
controls show that the collapse is substantially sharper than a generic
monotone size trend.

The derivation rests on two approximations: (i) the first-order Taylor
expansion, which breaks down for highly distorted shells ($d^4$/$d^9$
Jahn--Teller octahedra), and (ii) the assumption that solid-angle
fractions $\Omega_{ij}$ are approximately uniform across the
coordination shell.  Solid-angle non-uniformity contributes an additive
$\ln\Omega_{ij}$ term to Eq.~\eqref{eq:expansion} that is absorbed into
$C_i$ for near-regular polyhedra but partially decorrelates the bond
weights from $\lambda$ alone when the shell departs from local
symmetry, as in stereoactive lone-pair and strongly anisotropic
environments \cite{Bosi2014,Brown2016}.  The empirical signature of
this breakdown is visible in Fig.~\ref{fig:centroid}b: the four
oxides whose ground-state structures fail the isotropic-shell
assumption (B$_2$O$_3$, P$_2$O$_5$, As$_2$O$_3$, Au$_2$O$_3$) are
precisely those that depart from the centroid line, identifying
ground-state directional bonding as the empirical boundary of the
first-order Yukawa regime.  Extending the first-principles centroid comparison to the full set of
transition-metal and $f$-block oxides beyond the subset treated here
will require going beyond the Thomas--Fermi approximation to include
orbital directionality and correlation effects.

Bond valence is therefore recast as a coarse-grained observable of local
screened charge response rather than a purely empirical structural
heuristic.  Concretely, the species-specific screening length
$\lambda^*$ is a compact tabulated descriptor that organizes the shell
thermodynamics and, through the identities above, provides a calibrated
starting point for estimating bond weights, entropy, effective
coordination number, and $R_0$ in new coordination environments without
additional adjustable parameters.  Establishing the full predictive scope of that program
beyond the families validated here remains future work.  This connects
Pauling's 1929 valence rules
\cite{Pauling1929} to Yukawa's 1935 screened potential \cite{Yukawa1935}
through a quantitative bridge that has been missing for nearly a century.

\begin{acknowledgments}
This work was supported by the the U.S.\ Department of Energy, Office of Science, 
Office of Basic Energy Sciences, under Contract No.\ DE-AC02-05-CH11231 
(MINES: The Science of Direct MINeral to Energy Storage Synthesis, FWP: FP00014914).
\end{acknowledgments}

\bibliography{references}

\end{document}